\documentclass[fleqn,usenatbib]{mnras}
\pdfoutput=1
\usepackage{newtxtext,newtxmath}
\usepackage[T1]{fontenc}
\DeclareRobustCommand{\VAN}[3]{#2}
\let\VANthebibliography\thebibliography
\def\thebibliography{\DeclareRobustCommand{\VAN}[3]{##3}\VANthebibliography}
\usepackage{graphicx}
\usepackage{amsmath}
\usepackage{xcolor}
\usepackage{pdflscape}
\newcommand{\beq}{\begin{equation}}
\newcommand{\eeq}{\end{equation}}
\newcommand{\beqa}{\begin{eqnarray}}
\newcommand{\eeqa}{\end{eqnarray}}
\def\lap{\lower.5ex\hbox{$\; \buildrel < \over \sim \;$}}
\def\gap{\lower.5ex\hbox{$\; \buildrel > \over \sim \;$}}

\title[Flat Cosmic Patterns]{Flat Patterns in Cosmic Structure}

\author[P. J. E. Peebles]{P. J. E. Peebles$^{1}$\thanks{E-mail: pjep@Princeton.edu}\\
$^{1}$Joseph Henry Laboratories, Princeton University, Princeton, NJ 08544, USA\\
}

\pubyear{2023}

\begin{document}
\label{firstpage}
\pagerange{\pageref{firstpage}--\pageref{lastpage}}
\maketitle
\begin{abstract}
It is natural to wonder how far the flat pattern in the distribution of galaxies and clusters of galaxies around the de Vaucoueurs Local Supercluster extends, and whether there are other similarly extended flat patterns in the large-scale structure of the universe. I present evidence of two extended flat and thin sheet-like patterns in the distributions of galaxies and clusters detected at redshift $z<0.021$. Sheet~A contains our position and is tilted $11^\circ$ from the supergalactic pole, meaning the Local Supercluster is a moderately bent part of the more extended Sheet~A. The continuation of this sheet is detected in the disjoint sample of galaxies at redshifts $0.021<z<0.041$ and again in the disjoint samples of galaxies and clusters of galaxies at $0.042<z<0.085$. Sheet~B is 15~Mpc from us at its closest point. It is detected at $z<0.021$ and at $0.021<z<0.041$. These results make a serious case for the reality of signatures of close to flat and thin extended sheet-like patterns in cosmic structure, and an interesting challenge for the $\Lambda$CDM cosmology. 
\end{abstract}

\section{Introduction}

Vera Cooper Rubin (1951) and G\'erard de Vaucouleurs (1953, 1958) pointed out that the  nearby galaxies tend to be concentrated in a flat and maybe rotating system. I am not aware of detection of rotation, but the flattened distribution is clearly seen and variously known as the Virgo Supercluster, the Local Sheet (Tully, Shaya, Karachentsev, et al. 2008; McCall 2014) or, following de Vaucouleurs, the Local Supercluster. Two questions naturally arise. What is the detectible extent of this flat pattern with greater than average  number density? And are there other similarly extended and close to flat patterns in cosmic structure? Important steps to addressing the first question include the demonstrations by  Zel'dovich, Einasto, and Shandarin (1982, Fig. 5), Einasto, Corwin, Huchra,  Miller, Tarenghi (1983, Fig. 1), and Tully (1986, 1987) that clusters of galaxies tend to be close to the plane of  the Local Supercluster in a region several hundred megaparsecs across. Einasto, Tago, Jaaniste, Einasto, and Andernach (1997) remark that ``One example of [sheets or planes] is the Supergalactic Plane, which contains the Local Supercluster, the Coma Supercluster, the Pisces-Cetus and the Shapley superclusters.'' Of equal importance is the evidence presented by Shaver (1990, 1991) and Shaver and Pierre (1989) that radio galaxies at  distances $z\lap 0.02$ also tend to be at low supergalactic latitudes. Observations along these lines are not widely discussed but they are  informative and potentially important aspects of the large-scale nature of the universe. An analog on a smaller scale, the ``plane of satellites problem,'' is more widely discussed. Considerations along this line trace back at least to Lynden-Bell (1976), who pointed out interesting alignments of high-velocity HI clouds and dwarf satellites of the Milky Way. The length scales differ but the philosophy is the same: Ask what careful examination of the world around us might reveal that is new and maybe informative about the nature and origin of the universe. 

The examination of distributions of galaxies and clusters of galaxies presented here yields a  case for detection of two flat and thin regions of greater than average mean density, identified as exceptionally large counts of galaxies in disk-like cells 5~Mpc thick and 150 Mpc across. One sheet is close enough to the direction and position of the de Vaucouleurs Local Supercluster to justify naming it the Extended Local Supercluster. The sheet is detected in three disjoint saples of redshift, the furthest at redshift-based distance 180 to 360~Mpc. It undoubtedly is real by the test of reproducibility of detection in separate ranges of redshift, and it is an interesting challenge to the standard cosmology. 

The results presented here have in common the large length scales associated with great walls of galaxies (Huchra and Geller, 1982;  Gott, Juri{\'c}, Schlegel, et al. 2005; Lietzen, Tempel, Liivam{\"a}gi et al. 2016; and references therein) and Large Quasar Groups (Clowes, Harris, Raghunathan, et al. 2013; Friday, Clowes, Williger 2022, and references therein). The cosmic web is said to contain sheets of objects, but the sheets typically are curved. The sheets to be considered here are modeled after the Local Supercluster: thin, extended, and quite close to flat. And the method of detection is meant to be applied  equally well to observations and to numerical simulations of cosmic structure.

The redshifts of objects in this study are not large, $z \lap 0.1$, so I ignore space curvature and other relativistic effects and use the low distance conversion of redshift $z$ to distance $r = cz/H_{\rm o}$, with the Hubble parameter 
\beq
H_{\rm o} = 70 \hbox{ km s}^{-1}\hbox{ Mpc}^{-1}.\label{eq:Hnot}
\eeq
This is not always a safe approximation for nearby galaxies, but in the central part of the de~Vaucouleurs  Local Supercluster, at distances $\lap 10$~Mpc, many of the galaxies have measured or sensible inferences of distances. For objects that are further away distances based on redshifts seem adequate.

The probe or filter used in this study for the discovery of extended flat concentrations of galaxies is explained in Section~\ref{sec:probe}.  A test by its application to the distribution of galaxies in the Nearby Universe, at distances less than about 10~Mpc, is reported in Section~\ref{sec:LU}. The probe shows that the distribution of galaxies in the Local Supercluster is quite flat, as we already new. Section~\ref{sec:z02} presents results of applications of the probe to the distributions of galaxies and clusters of galaxies at distances less than redshift $z_{\rm max} =0.021$, or about 90~Mpc, a reasonable outer bound of the Local Universe. The application yields evidence of two similarly thin and flat sheet-like patterns in the spatial distributions of galaxies and clusters. Section~\ref{sec:conditions} presents results of exploration of the natures and contents of these two candidates for true flat sheets. The results of applications of the probe at greater distances, which reveal evidence of continuations of the two sheets, are presented in Section~\ref{sec:z08}. Concluding remarks are in Section~\ref{sec:discussion}. 
 
\begin{figure}
\begin{center}
\includegraphics[angle=0,width=2.5in]{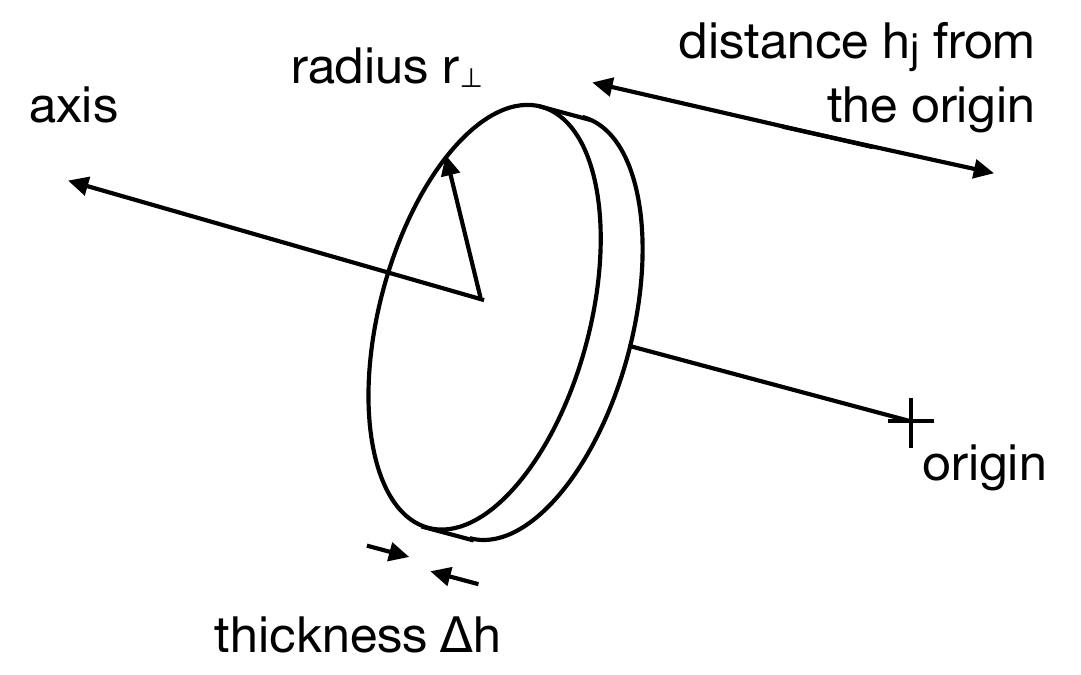} 
\caption{Objects are counted in layers of disk cells with radius $r_\perp$ and thickness $\Delta h$. The axis passes through the disk centers normal to the disks. The stack of these disks define a truncated cylinder. The positions $h_j$ of the disk midplane are measured from the origin, our position.}\label{fig:probe}
\end{center}
\end{figure}

 \section{A Probe for Planar Patterns}\label{sec:probe}

The probe, or filter, used to detect close to flat patterns in the distributions of objects is illustrated in Figure~\ref{fig:probe}. Objects are counted in stacks of cells, each cell a disk with radius $r_\perp$ and  thickness $\Delta h$. A stacked array of $2N+1$ adjacent disks centered on and normal to an axis passing through the origin defines a cylinder truncated by planes normal to the axis. The distance measured along the axis from the origin to the centerplane of the $j^{\rm th}$ disk is $h_j = j\times \Delta h$, with $j = 0, \pm 1, \pm2$, to $\pm N$. The origin is at the centerplane of the $j = 0$ cell. The two ends of the cylinder, at distance from the origin
\beq
h_{\rm max} = (N + 1/2)\Delta h,
\eeq
fit in a sphere of radius $R$. At maximum volume of the cylinder given the radius of the sphere the dimensions are 
\beq
R = \sqrt{3} h_{\rm max}, \quad r_\perp = \sqrt{2} h_{\rm max}.\label{eq:dimensions}
\eeq
The applications presented here use $N=10$ for 21 disk cells. The ratio of the disk cell diameter to the disk thickness is then 
\beq
2 r_\perp /\Delta h = 2^{3/2}(N+ 1/2) = 29.7.\label{eq:aspectratio}
\eeq
The large aspect ratio,  30:1, means the filter is sensitive to quite flat and thin patterns in cosmic structure. 

The intuition that this arrangement might be an interesting probe is drawn from the evidence that the Local Supercluster is quite flat and that radio galaxies and clusters of galaxies tend to be close to the extended plane of the Local Supercluster (e.g. Einasto et al.~1983; Tully 1987; Shaver 1990). As mentioned above this probe, or filter, is not sensitive to curved sheets of galaxies in great walls (e.g. Lietzen et al. 2016 and references therein), or to the configurations of Large Quasar Groups (e.g. Clowes, Harris, Raghunathan, et al. 2013 and references therein), unless these configurations are flat. 

The advantages of this filter are that the judgement of significance of detection is simplified by the rigidity of the definition, and that it would be easy to compare results of application of the filter to observations and to numerical simulations of cosmic structure in a theory such as $\Lambda$CDM. It must be noted, however, that despite the claim of rigidity I do not hesitate to postulate ``bends'' that offer intuitively reasonable continuations of sheets detected in disjoint ranges of redshift.

A convenient disc cell thickness for the search for flat patterns in the Local Universe  is $\Delta h = 5$~Mpc. With $N=10$ the separation of midplanes of the disks on each end of the cylinder is 100~Mpc, and the full length of the cylinder is $2h_{\rm max} = 105$~Mpc. The cylinder has radius $r_\perp = 74.2$~Mpc, which fits in a sphere of radius $90.9$~Mpc. This radius, with the Hubble parameter in equation~(\ref{eq:Hnot}) is redshift 
\beq
z_{\rm max}=0.0212  \label{eq:hnot}
\eeq
or $z_{\rm max}=0.021$ for short. This bound is a convenient limiting redshift of the Local Universe. The ratio of disk cell diameter to thickness, 30:1, is about right for the search for patterns as flat as the Local Supercluster. 
  
At given orientation of the axis of this cylindrical arrangement, objects are sampled only inside the cylinder, not all that are in the sphere. Different objects in the sphere enter the counts in disks in cylinders with different orientations. Data would be more completely sampled if the disks closer to the origin had larger radii, but trials indicate that taking account of the different disk areas is not a useful complication. More important for this exploratory survey is the uniformity of sampling by the disk cells. 

The search for close to plane patterns in the second redshift bin takes account of the smaller number densities of objects in the catalogs used here by doubling dimensions to redshift bounds $0.021<z<0.042$, disk cell thickness $\Delta h = 10$~Mpc., and cylinder radius $r_\perp = 149$~Mpc. This preserves the ratio 30:1 of cell diameter to thickness. We need disjoint samples and uniform sampling of the space covered by the disk cells. This is done by replacing the solid cylinder used for the Local Universe by a hollow cylinder, a straight tube. The inner radius of the tube is $r_\perp^{\rm min} = R/2$, which removes the data from the closer redshift bin. It follows from equation~(\ref{eq:dimensions}) that the inner radius of the tube is 
\beq
r_\perp^{\rm inner} = \sqrt{3/8}r_\perp. \label{eq:innerradius}
\eeq
The inner tube radius is 0.6124 times the outer tube radius.\footnote{In the first draft circulated on the arXiv I erroneously took the inner radius of the tube to be half the outer radius, not 0.61 times the outer radius. This allowed inclusion of  a small slice of data from the next nearer redshift bin. Correcting this error changed all figures in the second and third redshift bins in the first draft. The changes are small, mostly difficult to see, except in the fourth redshift bin, where the sample becomes too small for interesting analysis.} This means the volume sampled in the second redshift sample is five eights of the volume of the solid cylinder. 

The cell used for the third redshift bin,  $0.042<z<0.085$ doubles all dimensions again.

An alternative approach uses a cylinder of solid disk cells applied to positions of objects in the distance range $R/2$ to $R$. The problem with this is that the disk cells at distance less than half the sphere radius have smaller mean counts than in cells further out. Defining interestingly large counts is complicated in this situation. Again, the uniformity of disjoint sampling is preferable.

Candidates for planar patterns of larger than average density are discovered as exceptionally large counts found in directions of the cylinder or tube of disk cells chosen isotropically at random (to the extend that the random number generator can be trusted). The count, direction, and midplane distance $h_j=j\times \Delta h$ of the cell is recorded when the count in the cell exceeds any previously found. Perhaps the cell found this way contains a flat and thin sheet with systematically larger than average number density, but this procedure produces candidates for flat configurations even if there are none. The prime evidence of reality is the detection of continuations of sheets in disjoint ranges of redshift. This is demonstrated in examples to be presented.  

The direction of the axis that produces a large count in a disk cell, maybe an indication that the cell contains a flat sheet, is stated to the nearest degree (with an exception to be mentioned). This is because discreteness noise in the sample produces discrete directions at maximum cell count or close to it,  typically spread in direction over a degree or so. I have experimented with allowing the zero-point of $h$ to be shifted from the origin, thus broadening the search for maximum count. In trials this addition to the parameters to be scanned yields larger counts in scattered values of the shift and direction. The choice of the best shift requires specifying the direction more closely than a degree, however, which I doubt is meaningful. So the zero-point of $h$ is fixed at the origin.

By convention the axis of the cylinder of disks points to the north galactic hemisphere. This means the directions of interestingly large counts  that are close to each other but on opposite sides of the galactic equator appear far apart in a map of directions confined to the north hemisphere. The working solution is presented in Section~\ref{sec:2thinsheets}. 

The mean count of objects in disk $j$ averaged over random orientations is a measure of the mean number density at distances in the range from $h_j$ to $r_\perp$. The convention of north-pointing axes means the mean densities at $h_j$ and $-h_j$ usually differ. This convention makes the mean counts as functions of $h_j$ a useful indication of the degree of departures from homogeneity of a sample of objects, though one that would be difficult to translate to density as a function of spatial position.

The application of this filter with 21 disk cells and the largest sample considered here, $\sim 8000$ galaxies, using basic fortran~90 with no attempt to improve efficiency of computation, is repeated a million times in six minutes by my 5 year old desktop computer.
 
\begin{figure}
\begin{center}
\includegraphics[angle=0,width=2.5in]{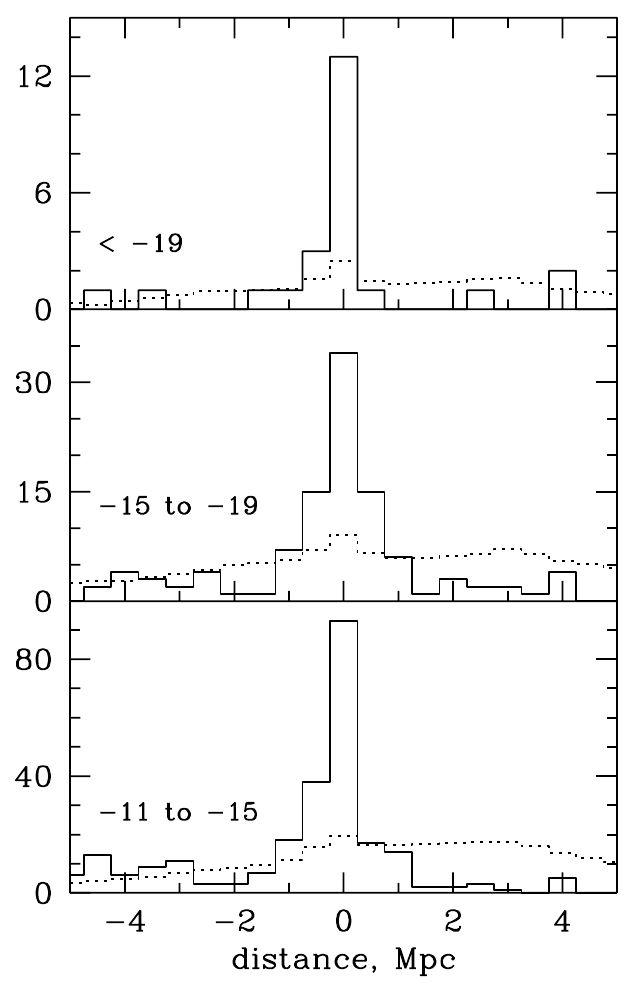} 
\caption{Counts of galaxies closer than 9.1~Mpc in disk cells 0.5~Mpc thick with radius 7.4~Mpc along the same axis in each range of luminosity. The dotted histograms show mean counts.}\label{fig:NU}
\end{center}
\end{figure}

\section{The Nearby Universe}\label{sec:LU}

The probe is tested by application to the positions of galaxies closer than about 10~Mpc, where the concentration of galaxies to the plane of the de~Vaucouleurs Local Supercluster is familiar. Figure~\ref{fig:NU} is based on the NASA NEARGALCAT or Nearby Universe (NU) version of the Karachentsev, Makarov, and Kaisina (2013) Updated Nearby Galaxy Catalog. 

The histograms in Figure~\ref{fig:NU} are based on counts of galaxies closer than 9.1~Mpc. At this distance the NU contains 53 galaxies more luminous than $M_B < -19$ and 243 NU galaxies with absolute magnitudes $-19 < M_B < -15$. The catalog contains 461 galaxies with $-15 < M_B < -11$, which I use, though I appreciate that many more of these faint galaxies likely are not yet cataloged. 

The counts are in disk cells 0.5~Mpc thick with diameter  $2r_\perp =  14.8$~Mpc. The cells are thirty times as broad as they are thick. This is a test for quite thin patterns.

Figure~\ref{fig:NU} shows counts as functions of distance $h_j = 0.25\times j$~Mpc  along an axis common to the three luminosity classes. The axis was chosen to get close to maxium counts in all three luminosity bounds. The normal to the disk cells is directed to galactic and supergalactic coordinates
\beq
l = 48^\circ,\ b = 1^\circ,\ SGL = 277^\circ,\ SGB = 85^\circ.\label{eq:LUdirections}
\eeq
As explained in Section~\ref{sec:probe} the angles are given to the nearest degree; shot noise makes more precise values meaningless. To this precision the counts in the central disk cell, $h_j=0$, are close to the largest counts found by adjusting the direction of the axis separately for each luminosity interval.

The direction in equation~(\ref{eq:LUdirections}) is tilted $\sim 5^\circ$ from the  supergalactic pole SGB = 90$^\circ$. When the axis is directed to SGB = 90$^\circ$ the peak count of galaxies is 8 of the most luminous ones, fewer than the 13 when the pole is in the direction in equation~(\ref{eq:LUdirections}). In the intermediate luminosity interval the peak count is 27 at SGB = 90$^\circ$, fewer than the 34 for the tilted pole. In the lowest luminosity interval there are 77 galaxies in the central cell when the axis directed to SGB = 90$^\circ$, again fewer than the 93 for equation~(\ref{eq:LUdirections}). The differences are modest but systematic. By this measure the galaxy distribution in the Nearby Universe is tilted by about five degrees from the plane that de~Vaucouleurs and colleagues established (as defined by the NASA/IPAC Coordinate Calculator). This is impressively close considering the far more limited data available then. 

The histograms plotted as dotted lines in Figure~\ref{fig:NU} are the averages over directions of the counts in the disk cells as a function of distance $h_j$ from the origin. By convention the axis points to the north galactic plane, and we see that there are more of each luminosity class in the galactic north. The peaks in the dotted histograms at $h_j=0$ might be caused in part by greater completeness at closest samples, and maybe in part by the prominent concentration of galaxies to low supergalactic latitudes. 

The conclusion from this application is that the probe for flat thin sheet-like patterns in cosmic structure yields a reasonable result  in a known situation. 

\section{The Local Universe}\label{sec:z02}

The Local Universe is taken to be the redshift range
\beq
0< z < 0.021.\label{eq:zbounds}
 \eeq
With the Hubble parameter in equation~(\ref{eq:Hnot}) this is distance less than about 91~Mpc. The upper bounds on the redshift and distance are not round numbers because they are determined by a convenient disk cell thickness for the search for traces of flat patterns in the distributions of objects in the Local Universe. The cell dimensions are 
\beq
 \Delta h =5\hbox{ Mpc},\ r_\perp =  74.25 \hbox{ Mpc},\ \hbox{diameter/thickness} \sim 30, \label{eq:dist_1_dimensions}
 \eeq
 for 21 disk cells.

\subsection{Two Sheets of Galaxies}\label{sec:2thinsheets}

The search for flat concentrations of objects is simplest to interpret when applied to samples that cover most of the sky. The main galaxy sample used here is the 2MRS catalog of galaxy positions and redshifts from the NASA TWOMASSRSC catalog of galaxies detected in the 2MASS survey selected by apparent magnitude $K_s < 11.75$ at $2.2\mu$ (Skrutskie, Cutri, Stiening, et al. 2006), with redshifts and morphological types from Huchra, Macri, Masters, et al. (2012). To reduce radial gradients in the number density in a sample with redshift bound $z < z_{\rm max}$ I use the apparent magnitude bound 
\beq
K_s < 11.75 + 5 \log_{10} (z/z_{\rm max}). \label{eq:Kband_bound}
 \eeq
 
The condition in equation~(\ref{eq:zbounds}) that the redshift is positive removes some Nearby Universe galaxies. Other NU galaxies are in this sample, with distances computed from redshifts, but they are few enough that the distance errors are not likely to influence the search for sheet-like patterns on distance scales an order of magnitude larger. 

I have not found evidence that obscuration in the plane of the Milky Way has interfered with the searches for flat patterns. The situation likely would different in the search for planar patterns of lower than average number density. 

\begin{figure}
\begin{center}
\includegraphics[angle=0,width=3.25in]{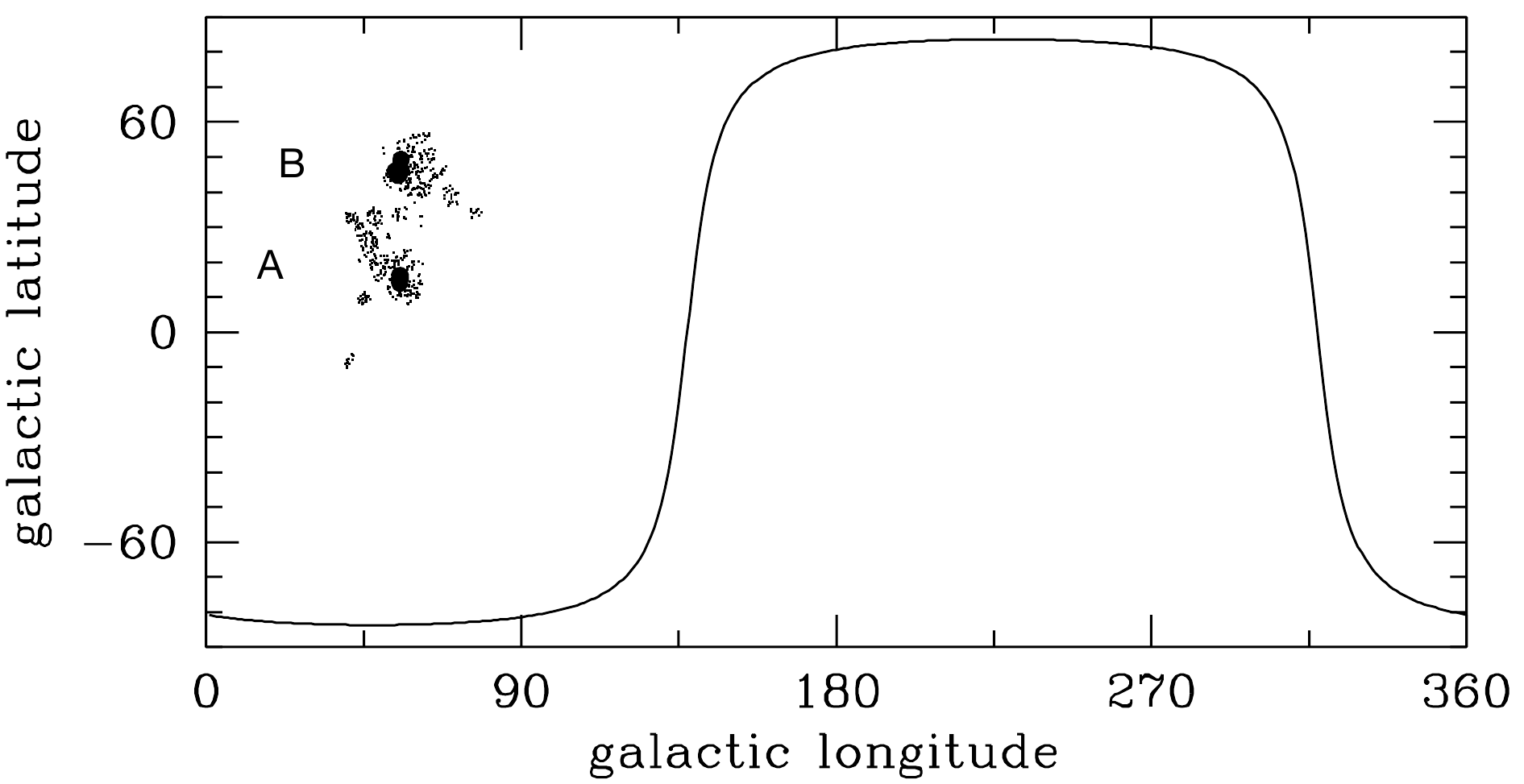} 
\caption{Directions of axes in whic h the count in a cell of 2MRS galaxies at $z<0.021$ exceeds 500, at the dots, and  600, at the filled black regions. A position below the curved line is equivalent to a position above the line with the distance $h$ reckoned in the opposite direction.}\label{fig:survey11}
\end{center}
\end{figure}

Figure~\ref{fig:survey11} shows the result of a search across the sky and at redshifts less than 0.021 for exceptionally large counts in cells with the dimensions in equation~(\ref{eq:dist_1_dimensions}). This is based on the 8540 2MRS galaxies at redshifts $0<z<0.021$. An angular position of the axis below the curved line in the figure is equivalent to a position above the line belonging to the same axis with the disk cell distances $h$ labeled in the opposite direction. The dividing line in the figure is $90^\circ$ from the de~Vaucouleurs supergalactic pole at $l=47.37^\circ$,  $b= 6.32^\circ$ in standard galactic coordinates. This arrangement was motivated by the small clump of points in the figure at negative galactic latitude and longitude similar to that of the two prominent clumps at positive latitude. An equivalent figure that shows all directions at positive latitudes puts the small clump at longitude about  $180^\circ$ away from the prominent clumps, which is undesirable.

In the figure directions of axes for which a count in a cell is 500 or more galaxies are plotted as dots, and the solid black areas show directions in which the count in a disk cell is 600 or more. (This is from $3\times 10^4$ randomly chosen directions. In larger samples the points merge, making it difficult to see the solid black areas that are closer to the maximum counts.)

The sky survey in Figure~\ref{fig:survey11} reveals two directions in which the counts of galaxies are particularly large in one of the thin cells in the distance range $-10\leq j \leq 10$. The fiducial normals in the directions at which the cell counts are close to maximum, and the ranges of distance from the origin contained by the cells with the maximum counts, are
\beqa
&&\hbox{ Sheet A$_1$:}\, l = 55^\circ, \, b = 14^\circ,\, -2.5<h<2.5\hbox{ Mpc}, \nonumber \\
&&\hbox{ Sheet B$_1$:}\, l = 55^\circ,\, b = 46^\circ, \,  -17.5<h<-12.5\hbox{ Mpc}.
 \label{eq:AB}
\eeqa
The subscripts are meant to indicate that these are based on the data in the first redshift sample, or bin. 

The small group of dots in Figure~\ref{fig:survey11} at similar longitude and negative latitude show a third candidate sheet, but one with maximum cell counts of about 500 galaxies, while the maximum counts in the two named sheets are above 600. By this measure Sheets~A$_1$ and~B$_1$ are uniquely prominent in the Local Universe, at $z<0.021$. Bear in mind that these are based on counts in thin disk cells, aspect ratio 30:1. 

\begin{figure}
\begin{center}
\includegraphics[angle=0,width=2.75in]{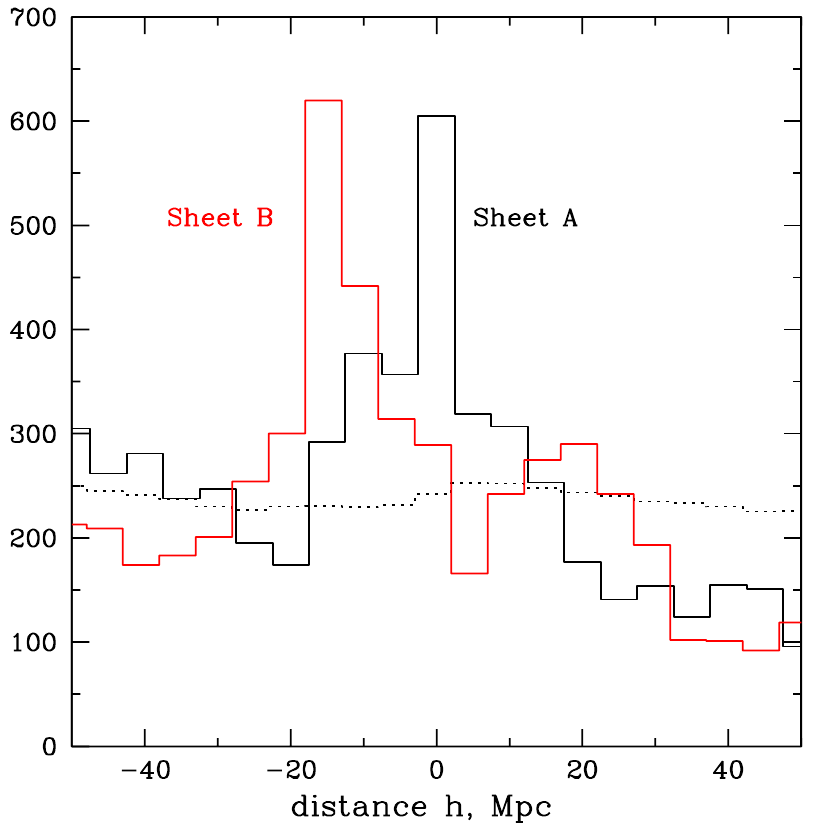} 
\caption{Counts of 2MRS galaxies at redshift $z < 0.021$ in disk cells as functions of distance along the axis in the direction of the normal of Sheet~A$_1$, plotted in black, and of Sheet~B$_1$, plotted in red. The dashed histogram shows counts averaged over orientation.}\label{fig:2sheets}
\end{center}
\end{figure}

Figure~\ref{fig:2sheets} shows counts of  2MRS galaxies closer than $z=0.021$ in disk cells with normals in the directions in equation (\ref{eq:AB}), as functions of the distance $h_j$ of the cell centerface from the origin. The histogram plotted as the solid black line shows counts along the fiducial axis of Sheet~A$_1$, and the histogram plotted in red shows counts along the direction of Sheet~B$_1$. The histograms look similar, with pronounced peaks. The maximum count in the cell that contains Sheet~A$_1$  is 605 galaxies and the counts in the two adjacent cells, 5~Mpc away, are 357 and 319. In Sheet~B$_1$ the maximum count is 620, and the counts in the adjacent cells are 300 and 442. It is notable that the galaxy counts at the two maxima are in cells that are just 5~Mpc thick and thirty times that wide, and that the counts are considerably smaller in the neighbouring cells centered just 5~Mpc away.

The normal of Sheet~A$_1$ is $15^\circ$ from the normal of the Nearby Universe in equation~(\ref{eq:LUdirections}), and $11^\circ$ from the de Vaucouleurs supergalactic pole. When the axis is directed to the supergalactic pole the peak count, still at $j=0$, is 444, appreciably down from 605 for Sheet~A$_1$. This is evidence of reality of the tilt of Sheet~A$_1$ from the supergalactic pole, by a modest amount. 

The planes of the Nearby Universe and Sheet~A$_1$ both pass through the origin, our position, in directions that differ by just $15^\circ$. This means the Local Supercluster can be taken to be a moderately bent part of the more extended Sheet~A$_1$, a configuration that can be termed the Extended Local Supercluster. This is not new. Einasto et al. (1983) found the alignment of clusters with the plane of the Local Supercluster at similar distances, and Shaver (1990) demonstrated a similar alignment of radio galaxies. But the greater numbers of 2MRS galaxies with measured redshifts allows a more accurate demonstration of a remarkably broad, flat, and thin region in which galaxies tend to be distinctly more numerous than elsewhere. 

When the axis of the disk cells is aligned with the normal of Sheet~B$_1$ the maximum count (in the red histogram in Fig.~\ref{fig:2sheets}) is at distance $j = -3$. The negative value means that at its closest Sheet~B$_1$ is in the cell that is centred 15~Mpc from us in the direction $l = 235^\circ$, $b=-46^\circ$. This axis is tilted $40^\circ$ from the de Vaucouleurs pole. The larger angle and the significant distance to Sheet~B$_1$ make it a candidate for a different flat sheet in which galaxies also tend to be unusually common in a broad thin pattern.

We must pause to consider the effect of distortions of distances by proper motions. Sheet~A$_1$ passes through our position, so distorted distances derived from redshifts do not tend to move objects into or out of the disk cell that contains our position. Sheet~B$_1$ is 15~Mpc away, and in the direction where the sheet is closest random peculiar motions $\sim 300$~km~s$^{-1}$ would produce distance errors $\sim 4$~Mpc. If uncorrelated they would tend to move objects out of the 5~Mpc thick disk cell, yet the two peaks in the histograms in Figure~\ref{fig:2sheets} are similarly narrow. But in the disk cell at distance $h\sim -15$~Mpc the effect is largest only toward the center of the cell, which is much broader than 15~Mpc. In more distant samples to be discussed the disk thicknesses are larger and less likely to be influenced by peculiar motions. I conclude that peculiar motions are not likely to have affected this search for flat patterns.  

\begin{figure}
\begin{center}
\includegraphics[angle=0,width=2.75in]{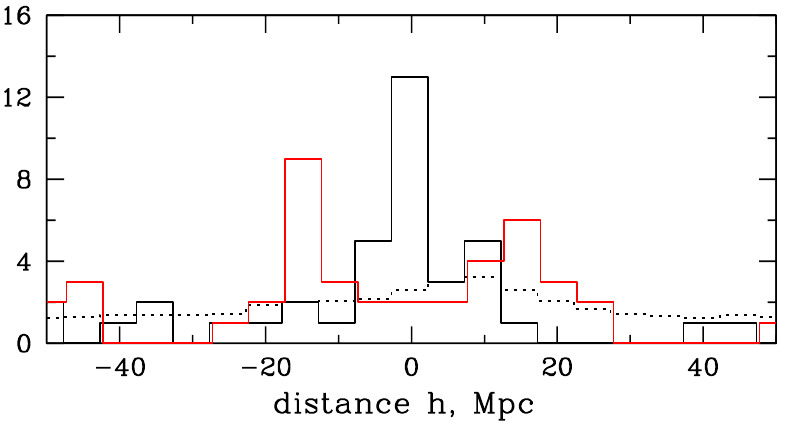} 
\caption{Counts of X-ray clusters of galaxies at $z<0.021$ measured along the direction of the cell that contains Sheet~A$_1$, plotted in black, and Sheet~B$_1$, plotted in red. The directions of the axes are set by the 2MRS galaxy sample. The dotted histogram shows the means, as in Fig.~\ref{fig:2sheets}.}\label{fig:CLs}
\end{center}
\end{figure}

\subsection{Reproducibility Checks}\label{sec:reproducibility}

The evidence of thin and flat sheet-like patterns is curious enough to call for a check of reproducibility. We have the observations that clusters are concentrated toward the plane of the Local Supercluster (Einasto et al.~1983; Tully 1986, 1987). An addition to this evidence is obtained from the cluster positions and redshifts compiled in the NASA MCXC Meta-Catalog of X-Ray Detected Clusters of Galaxies. There are not large numbers of clusters in this catalog, just 51 at  $z<0.021$, so I do not apply a bound on the X-ray flux densities.

Figure~\ref{fig:CLs} shows cluster counts in the 5 Mpc thick disk cells as functions of distance along the two orientations in equation~(\ref{eq:AB}), orientations derived from the galaxy distribution. The colour choice is the same as as in Figure~(\ref{fig:2sheets}). The peak counts are  reasonably distinct, similar, and at the same distances $h$ as in the galaxy counts in Figure~\ref{fig:2sheets}. It might be objected that this adds nothing new because cluster positions are known to be tightly correlated with galaxies. But the two thin and extended sheets are not intuitively expected in the standard $\Lambda$CDM theory, so it is important to have good evidence that the sheets are not statistical accidents or systematic errors in the measurement of a spatially homogeneous random process without true sheets.  Figure~\ref{fig:CLs} is an important addition to the case that thin, flat, and broad sheets of excess counts are real features of cosmic structure.

\begin{figure}
\begin{center}
\includegraphics[angle=0,width=2.75in]{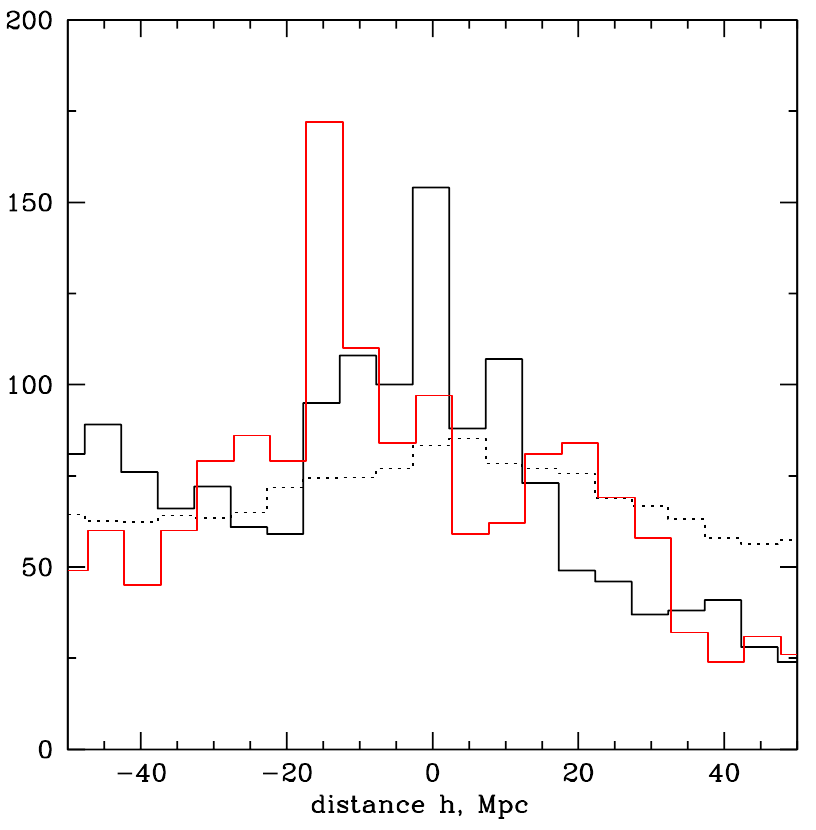} 
\caption{Counts of PSCz galaxies at $z<0.021$ along the directions of Sheets~A$_1$ (the black histogram) and~B$_1$ (the red histogram).}\label{fig:PSCz}
\end{center}
\end{figure}

The check in Figure~\ref{fig:PSCz} uses the PSCz redshift catalog that Saunders, Sutherland, Maddox, et al. (2000) drew from point-like sources in the IRAS (infrared astronomical satellite) sky survey at wavelengths from 12 to $100\mu$. The data used here were downloaded from the NASA HEASARC IRASPSCZ catalog, class GALAXY. The 2332 galaxies used to obtain Figure~\ref{fig:PSCz} have $60\,\mu$ flux densities bounded by
\beq
f  > 0.6(z_{\rm max}/z)^2 \hbox{ Jy}, \ z_{\rm max} = 0.021. \label{PSCzbound}
\eeq
This is a smaller sample than the 8540 2MRS galaxies, and the number density gradients indicated by the dotted line in Figure~\ref{fig:PSCz} are larger than in Figure~\ref{fig:2sheets}, but the peak counts in the directions and distances $h$ established from the 2MRS sample are clearly seen. At greater redshifts the PSCz galaxies reveal no evidence of Sheets~A or~B, maybe because the samples are not large enough.

These three samples at $z<0.021$ were obtained in different ways: 2MRS galaxies drawn from the 2MASS sky survey at $2\mu$ wavelength, PSCz galaxies identified as point-like sources in the IRAS sky survey at wavelengths $\sim 60\mu$, and clusters identified by detections of  X-ray emission by intracluster plasma. The consistent signatures of flat large-scale patterns at redshifts $z<0.021$ makes a good case for their reality.

\subsection{Probes of conditions in Sheets A and B}\label{sec:conditions}

Are Sheets~A$_1$ and~B$_1$ useful approximations to physically real flat thin patterns? If so do the galaxies in sheets differ in natures and distributions from objects outside the sheets, apart from the larger mean number densities? Preliminary considerations are discussed here. 
 
\begin{figure}
\begin{center}
\includegraphics[angle=0,width=3.in]{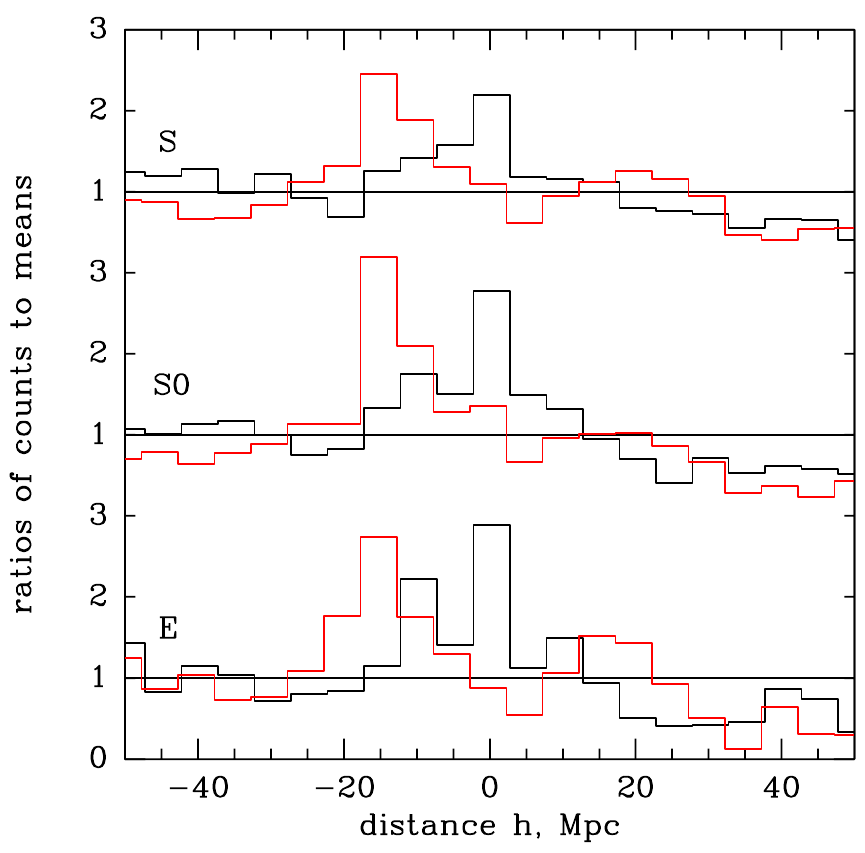} 
\caption{Ratios of the counts of 2MRS galaxies in disk cells at $z < 0.021$ to the mean count in that cell, separately for three morphological types, plotted in black for Sheet~A$_1$ and red for Sheet~B$_1$.}\label{fig:morphologyratios}
\end{center}
\end{figure}

The distributions of the most luminous of the elliptical and spiral galaxies in the 2MRS catalog relative to the plane of the Local Supercluster are compared in Peebles (2022). Figure~\ref{fig:morphologyratios} supplements this by comparing the distributions of counts in cells of the far more numerous 2MRS galaxies within the absolute magnitude limit in equation~(\ref{eq:Kband_bound}) at $z_{\rm max}=0.021$. The galaxies are labeled spiral, S0, or elliptical according to the prescription in Peebles (2022) for sorting the Huchra classification into these three classes.  At the cutoffs in redshift and luminosity there are 932 ellipticals, 2509 S0s, and 4105 spirals. The other 994 2MRS galaxies in the sample limited only by distance and luminosity, and used in Figures~\ref{fig:survey11} and~\ref{fig:2sheets}, have undetermined or enigmatic Huchra classifications. 

The effect of the different numbers of the three galaxy types is reduced in Figure~\ref{fig:morphologyratios} by plotting ratios of counts in a disk cell to the count in that cell at fixed distance and averaged over $10^6$ random orientations of the cylinder axis (as discussed in Sec.~\ref{sec:probe}). Sheet~A$_1$ is particularly interesting because it is a good candidate for the Extended Local Supercluster, but within fluctuations the ratios of counts to the means as functions of morphology and distance $h$ along the axis are similar for Sheets~A$_1$ and~B$_1$ (plotted in black and red). By this measure the two sheets are equally interesting. In both sheets the elliptical and S0 galaxies are more strongly concentrated to the sheets than are the spirals. The difference is much more modest than for the most luminous spirals and ellipticals (Peebles 2022), and it  agrees with the known tendency of the positions of ellipticals of more common luminosities to be more strongly correlated than the positions of spirals (Davis and Geller 1976). The slight and expected greater concentrations of ellipticals and S0s to the two sheets  suggest that there is nothing special about the galaxies in sheets, though a more detailed examination would be interesting. 

\begin{figure*}
\begin{center}
\includegraphics[angle=0,width=5.0in]{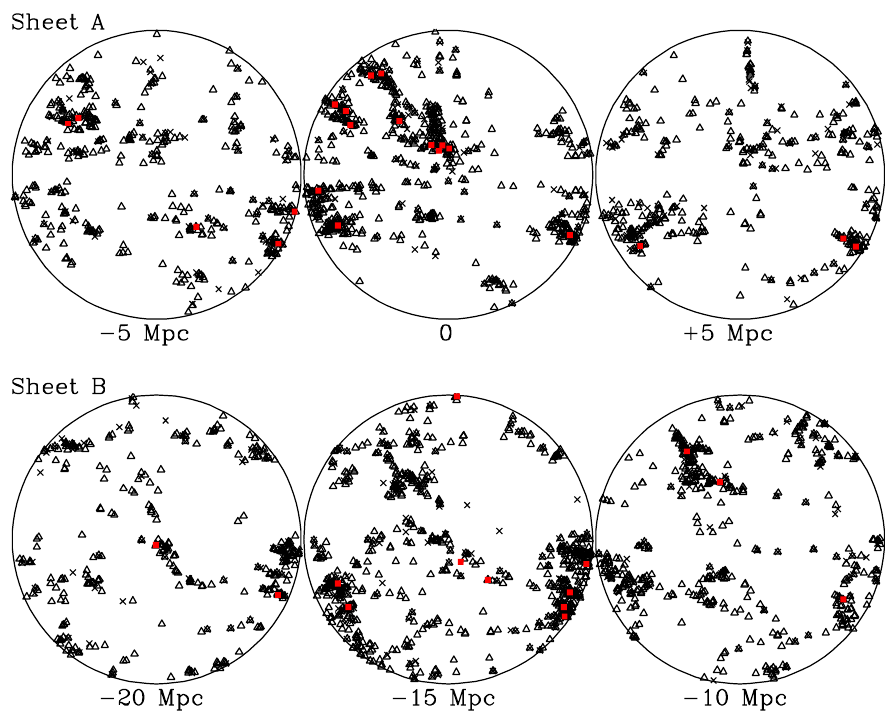} 
\caption{Maps of positions of 2MRS galaxies at $z<0.021$, plotted as black triangles, and clusters, plotted as red squares, viewed normal to the thin disks. The cells in the top row are aligned with Sheet~A$_1$, and aligned with Sheet~B$_1$ in the bottom row. The two centre maps show the distributions of objects in the disks with the largest counts. The maps on each side show positions in the disks on each side, at midplane distances $h_j$ from the origin.}\label{fig:maps}
\end{center}
\end{figure*}

A related issue is whether the nature of the space distribution of galaxies in sheets is unusual apart from the greater mean number density.  Figure~\ref{fig:maps} shows examples of the distributions of objects across the face of a disk cell 5~Mpc thick with radius 74.25~Mpc. The objects are selected by the redshift and luminosity bounds for Figure~\ref{fig:2sheets}. The distance of an object from the centre of each map is the perpendicular distance $r\sin(\psi)$ from the axis, which is the normal of Sheet~A$_1$ or~B$_1$, where $r$ is the redshift distance and $\psi$ is the angular distance of the object from the fiducial normal. The angular position $\mu$ is the azimuthal position of the object around the fiducial axis, with $\mu = 0$ when the object is in line with the fiducial pole and the North galactic pole. The horizontal and vertical axes in each map are $x = r\sin(\psi)\cos(\mu)$ and $y = r\sin(\psi)\sin(\mu)$. Positions are projected on the face of the disk cell, clusters marked by filled red squares, 2MRS galaxies by triangles. 

The centre maps in both rows show positions on the face of the disk cell that contains the largest count of  galaxies and, presumably, the thin flat sheet. The maps to the left and right show positions of objects in the two immediately adjacent disks centered 5~Mpc away. The considerably larger counts in the two central maps are evident, and expected from the prominent peaks in the two histograms in Figure~\ref{fig:2sheets}. The evidence the figure adds is that the greater numbers of objects in the central disk cells are not in orderly or otherwise unusual-looking arrangements. On and off the peak we see the familiar clumpy distributions of objects on scales ranging up to the distance across each of these maps. The simple difference is that number densities in the central disks  are larger than in the neighboring disks. The situation looks similar in and around Sheets~A$_1$ and~B$_1$.

\begin{figure}
\begin{center}
\includegraphics[angle=0,width=2.5in]{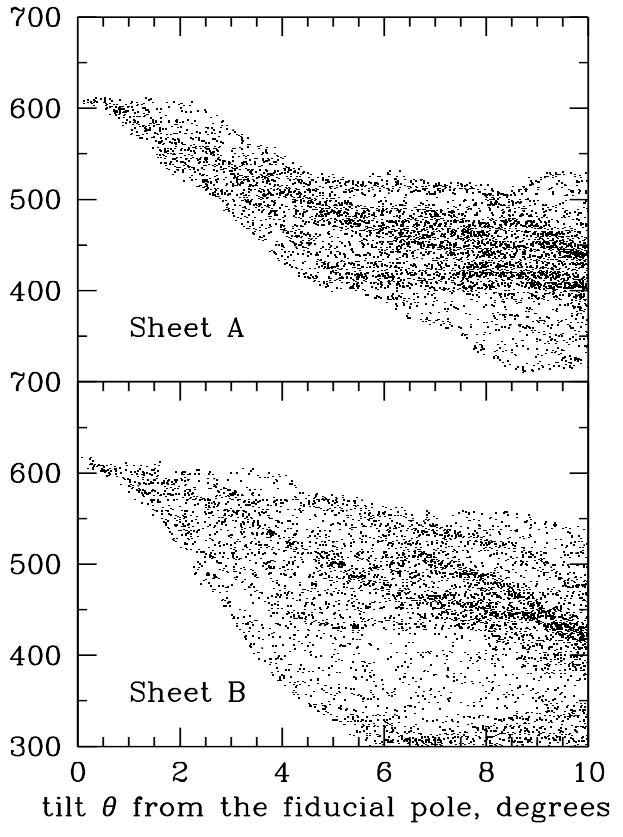} 
\caption{Variation of the galaxy count in a disk cell as the cell is tilted from the direction of maximum count.}\label{fig:shapes}
\end{center}
\end{figure}

Might the counts in a cell with the special orientations shown in Figure~\ref{fig:survey11} be exceptionally large because the cell contains a single particularly rich concentrations of galaxies? That is not my impression of the distributions of galaxies and clusters in the central maps in Figure~\ref{fig:maps}. And a single concentration would not produce the well-defined direction shown in Figure~\ref{fig:survey11}.  If the count in a thin disk cell were particularly large because the cell contains a straight filament of objects then there would be large counts in disk cells in other orientations  that also contain the filament. This is not suggested by the concentrations of points in the two directions in Figure~\ref{fig:survey11} . A filament of galaxies that is curved in two dimensions but flat in the third would serve as what I term a thin flat sheet.

An elaborated version of these thoughts starts from the consideration that the disk cells are much thinner than they are broad. The cell thickness $\Delta h$ divided by the distance to the edge of the disk cell, $r_\perp = \sqrt{2}(N + 1/2)\Delta h$, defines the characteristic angle, 
\beq
\Theta = [\sqrt{2}(N + 1/2)]^{-1} \sim 4^\circ.
\eeq
If the sheet of galaxies is scattered across the face of the disk, in the clumpy fashion suggested by the maps in Figure~\ref{fig:maps}, then a tilt  $\theta\sim\Theta$ from the orientation at maximum count in a cell, while the distance $h_j$ from the origin is fixed, moves the edge of the disk cell outside the original cell, meaning an appreciable part of the tilted cell is outside the nominal thin sheet. The expected size of the effect is difficult to estimate because we do not know how thick the sheet is, and the effect of the tilt depends on the abundance of objects just outside the cell before it was tilted. But a perceptible decrease in the count at tilt $\theta\sim\Theta$ is expected of an interesting approximation to extra galaxies in a flat sheet.

The data for the two scatter plots in Figure \ref{fig:shapes} are the 8540 2MRS galaxies that were used for Figures~\ref{fig:survey11} and~\ref{fig:2sheets}. The upper plot shows counts in cells centered on our position, $h_0 = 0$, as a function of the tilt $\theta$ from the normal of Sheet~A$_1$ (in eq.~\ref{eq:AB}) in random directions. The lower plot shows counts in cells centered at distance $h = -15$~Mpc as a function of the tilt from Sheet~B$_1$ in random directions. 

In this experiment the cell at $j=0$ is tilted while the centre of the cell is fixed. The cell at $j = -3$ is both tilted and shifted. Perhaps the scatter in counts at given tilt is greater around Sheet~B$_1$ in part because of this shift as well as tilt. But in both cases the counts at $\theta\sim\Theta\sim 4^\circ$ are about 80~percent of the counts at $\theta\sim 0$, roughly what one might expect of a thin sheet containing an unusually large count spread reasonably well across the diameter of the disk cell. By this measure Sheets~A$_1$ and~B$_1$ have the properties of thin flat regions in which the density is on average larger than usual across the width of the region sampled by the cell.

\section{More Distant Sheets}\label{sec:z08}

The smaller number densities at greater distances in the catalogs used here are compensated by doubling lengths in the second redshift bin, meaning redshifts $z = 0.021$ to 0.042, 10 Mpc thick disk cells with radius~$r_\perp=148.49$~Mpc. The data are kept disjoint by the use of a straight hollow cylinder, or tube, of thin disk cells with the inner radius given by equation~(\ref{eq:innerradius}) (in the form of a machinist's washer). That eliminates objects in the smaller redshift bin, as wanted. It reduces the use of data, but that is compensated by the advantage of disjoint samples scanned in an isotropic way. These dimensions are doubled again in the third redshift bin $0.042<z<0.085$.

\begin{figure}
\begin{center}
\includegraphics[angle=0,width=3.in]{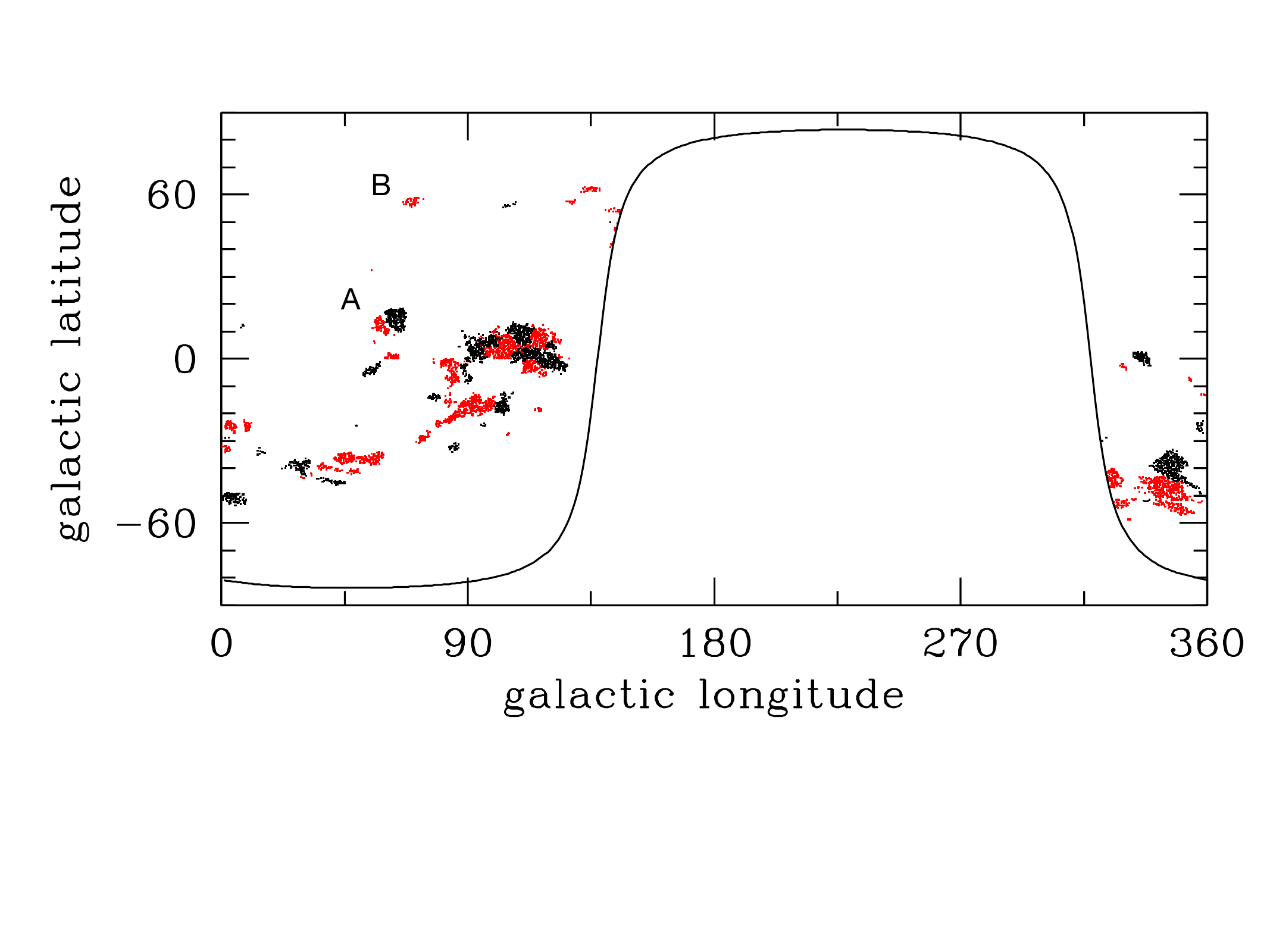} 
\caption{Evidence of continuations of Sheets~A$_1$ and~B$_1$ in the second redshift bin, $0.021<z<0.042$. Black points show directions where the galaxy count in the central cell, $j=0$, exceeds 220, and the red points show where the count at $j=-1$ or $-2$ exceeds 220. A direction below the curved line is the same as a direction above the line and in the opposite direction.} \label{fig:survey12}
\end{center}
\end{figure}

\subsection{Sheets A and B at Redshifts 0.021 to 0.042}

Figure~\ref{fig:survey12} shows evidence of detection of the continuations of Sheets~A$_1$ and~B$_1$ from the first redshift bin to the second, $0.021<z<0.042$. A threshold count of 220 galaxies in a hollow disk cell makes the continuations visible, but cell counts in cells in other orientations also exceed the threshold. The confusion is reduced by using the prediction that a continuation of Sheet~A$_1$ would contain the origin, $h\sim 0$, and a continuation of Sheet~B$_1$ would be at $h\sim -15$~Mpc at its closest. Accordingly, the black dots in the figure show orientations in which the galaxy counts exceed 220 in cells at the origin, $-5<h<5$~Mpc, and the red points show where counts are this large in one of the two cells that span the range of distances from the origin $-25<h<-5$~Mpc. The continuations of Sheets~A$_1$ and~B$_1$ at close to the predicted directions are labeled in Figure~\ref{fig:survey12}. The angular positions and distances from the origin at close to maximum cell counts of the two labeled clumps of points are 
 \beqa
&&\hbox{ Sheet A$_2$:}\, l = 63^\circ, \, b = 16^\circ,\, -5<h<5\hbox{ Mpc};\nonumber \\
&&\hbox{ Sheet B$_2$:}\, l = 62^\circ,\, b = 60^\circ, \,  -25<h<-5\hbox{ Mpc}.
 \label{eq:AB2}
\eeqa
The subscripts indicate that these are based on the data in the second redshift bin. 

The red points near A in Figure~\ref{fig:survey12} represent large counts in cells with normals near Sheet A$_2$ and further than 5~Mpc from the origin. The interesting concentration of disk cells with large galaxy counts and normals near $l\sim 100^\circ$, $b\sim 0$ seems to continue to smaller longitudes and lower latitudes  to $l\sim 300^\circ$, $b\sim -40^\circ$. I have not attempted to understand what this might mean. 

\begin{figure}
\begin{center}
\includegraphics[angle=0,width=3.in]{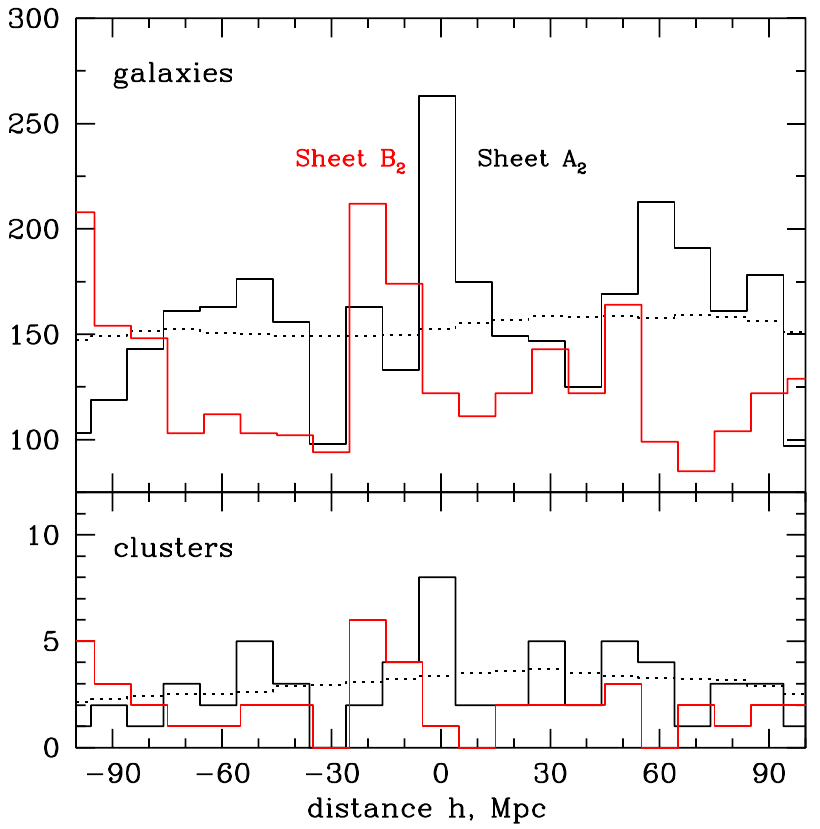} 
\caption{Counts of galaxies in the upper panel and clusters in the lower at $0.021<z<0.042$ along the directions of Sheets~A$_2$ (black) and~B$_2$ (red) in equation~(\ref{eq:AB2}). Dotted histograms are means.}   \label{fig:d2}
\end{center}
\end{figure}

Figure~\ref{fig:d2} shows counts of galaxies and clusters of galaxies in the rows of cells in the two directions in equation~(\ref{eq:AB2}). In the black histogram the disk cell centred on the origin contains Sheet~A$_2$, and in the red histogram the cell at $j = -2$ contains Sheet~B$_2$. As before the dotted histogram is the cell count averaged over direction as a function of distance $h$ of the disk cell from the origin. 

The fluctuations of counts around the means are large, but the two peak counts seem significant. The peak of Sheet~A$_2$ is in the distance interval $-5<h<5$~Mpc, consistent with the interval $-2.5<h<2.5$~Mpc for Sheet~A$_1$. The normal is tilted $8^\circ$ from Sheet~A$_1$. Sheet~B$_2$ peaks in the interval $-25<h<15$~Mpc, consistent with the interval $-17.5<h<-12.5$~Mpc in the first redshift bin. Its normal is $15^\circ$ from Sheet~B$_1$. 

The lower panel in Figure~\ref{fig:d2} shows counts of clusters along the orientations of the galaxy counts in the upper panel. It might be taken as somewhat encouraging that the two largest counts are at the expected distances from the origin, $h\sim 0$ and $h\sim -15$~Mpc. But there are other peak counts that are not much smaller, so this is a modest addition to the evidence of extensions of Sheets~A and~B to the second redshift bin. 

The case for detection of the extensions of Sheets~A and~B in the 2MRS galaxy distribution rests on the similar directions labeled A and B in Figure~\ref{fig:survey12}, along with the similar measures of the distances $h$ of the peak counts in Figure~\ref{fig:d2}. This passes the test of predictions from the data in the first redshift bin. It is important evidence of reality of the close to flat sheet-like patterns.

\begin{figure}
\begin{center}
\includegraphics[angle=0,width=3.in]{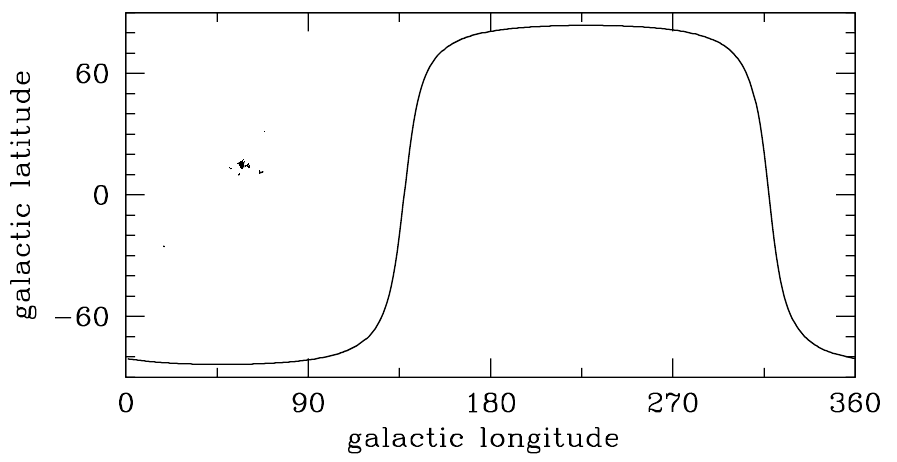} 
\caption{Evidence of continuation of Sheet~A to the redshift bin $0.042<z<0.0845$. The points show directions where the galaxy count in a hollow disk cell is $\geq 25$ at distances from the origin $-30<h<-10$~Mpc.} \label{fig:survey13}
\end{center}
\end{figure}

\subsection{Sheet A at Redshifts 0.042 to 0.085}

The counts in this third redshift bin are in a hollow cylinder or tube of disk cells 20~Mpc thick with inner and outer radii 182 and 297~Mpc, where the lower bound eliminates data in the second redshift bin. A survey of the sky in this redshift range and allowing the full range of distances $h$ from the origin reveals many directions where cell counts of galaxies reach 25, a relatively large value for the sample. As for the second bin this means the survey for Sheets~A and~B must be confined to a narrow range of distances $h$. Figure~\ref{fig:survey13} shows directions where the count of galaxies in a cell is 25 or larger in the disk cell at $ j=-1$, which spans the distance range $-30< h< -10$~Mpc. The choice $j=-1$ was empirical: it shows the most prominent concentration of directions near Sheet~A with large galaxy counts. 

I have not found evidence of the presence of Sheet~B in this redshift range and in trial scans of the sky in tubes at distance $h$ in the range $-4\leq j\leq1$.

\begin{figure}
\begin{center}
\includegraphics[angle=0,width=3.in]{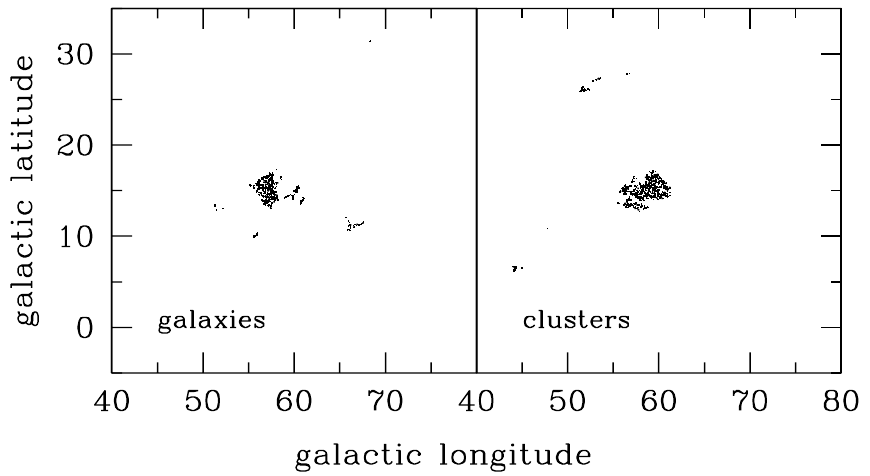} 
\caption{Evidence of continuation of Sheet~A to $0.042<z<0.0845$ from the distributions of galaxies and clusters. The points show directions where the galaxy count is $\geq 25$ or the cluster count is $\geq 30$ at distances $-30<h<-10$~Mpc.} \label{fig:survey123}
\end{center}
\end{figure}
 
Figure~\ref{fig:survey123} shows an enlarged region of the sky near the direction of the candidate for continuation of Sheet~A. The points are directions where the count of galaxies in a cell is 25 or larger, in the left-hand figure, and where the count of clusters is 30 or larger, on the right, again with $-30< h< -10$~Mpc.  Counts of galaxies and clusters both are close to maximum in the direction\footnote{As remarked in Sec.~\ref{sec:probe} it is meaningless to specify directions at maximum count to better than one degree, but I yield to the temptation to specify the galactic longitude to a half degree because this happens to bring both counts of galaxies and clusters close to maximum.} 
\beq
\hbox{ Sheet A$_3$:}\, l = 57.5^\circ, \, b = 15^\circ,\, -30<h<-10\hbox{ Mpc}.
 \label{eq:123}
\eeq

\begin{figure}
\begin{center}
\includegraphics[angle=0,width=2.75in]{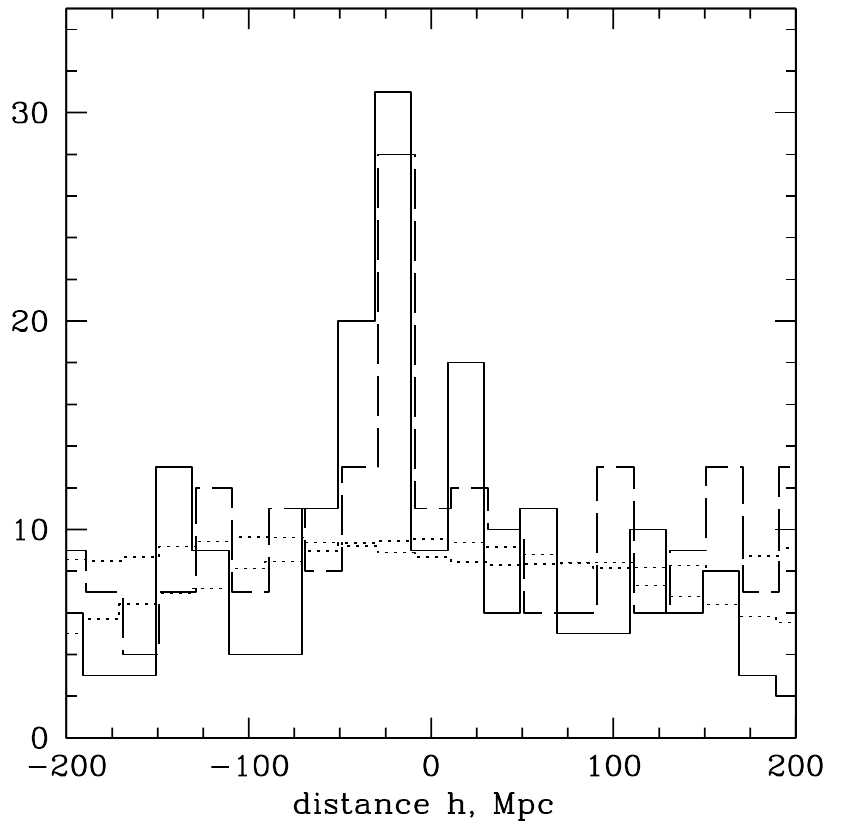} 
\caption{Counts of galaxies (dashed line) and clusters (solid line) at $0.042<z<0.085$ along the axis in equation~(\ref{eq:123}).}\label{fig:hist123}
\end{center}
\end{figure}

Figure~\ref{fig:hist123} shows counts of galaxies  and clusters of galaxies as functions of the distance $h$ of the disk cell from the origin along the direction in equation~(\ref{eq:123}). In this direction the disk cell at $h=-1$ contains Sheet~A$_3$. The counts of clusters are plotted as the solid line, and counts of galaxies as the dashed line. The dotted lines are the means, where the means for galaxies are larger than the means for clusters at the left-hand side of the figure.

At the redshift upper limit, $z_{\rm max}=0.085$, the absolute magnitude bound in equation~(\ref{eq:Kband_bound}) selects only the most luminous galaxies, which likely are in rich groups or clusters, though not necessarily the clusters selected by their X-ray emission. It is not surprising therefore that clusters and these luminous galaxies point to similar directions in Figure~\ref{fig:survey123}. But I emphasize once again that this agreement is important because it is a check that the small numbers of objects in the cells have not allowed statistical fluctuations to point to a meaningless direction. 

It is important also that the normal of Sheet~A$_3$ in equation~(\ref{eq:123}) is tilted just $5^\circ$ from the normal of Sheet~A$_2$ in the second redshift bin (eq.~[\ref{eq:AB2}]), $3^\circ$ from Sheet~A$_1$ in the first bin (eq.~[\ref{eq:AB}]), $17^\circ$ from the plane of the Nearby Universe (eq.~\ref{eq:LUdirections}), and $13^\circ$ from the supergalactic pole. The sheet of the Nearby Universe with Sheet~A$_1$ and Sheet~A$_2$ all pass close to the origin, at our position. The midplane of the disk cell that contains Sheet~A$_3$ is 20~Mpc from the origin (Fig.~\ref{fig:hist123}). This means the identification of the feature in Figure~\ref{fig:survey123} as a continuation of Sheet~A requires the postulate of a bend or warp. The warp is modest compared to the 600~Mpc diameter of the disk cells, and it is in the spirit of acceptance of the modest bends in directions required to identify continuations of the sheets across disjoint redshift bins. 

I conclude that the measures in the bin $0.042<z<0.085$ provide a serious addition to the evidence of the presence of Sheet~A.

\begin{figure}
\begin{center}
\includegraphics[angle=0,width=2.5in]{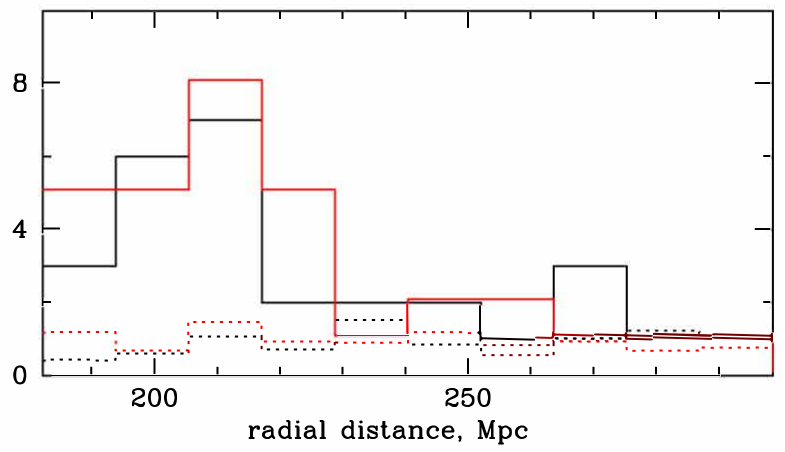} 
\caption{Distributions of radial distances of the galaxies, black, and clusters, red, that are in the disk cell $j = -1$ that contains Sheet~A$_3$. The dotted histograms are counts averaged over directions.} \label{fig:disthist}
\end{center}
\end{figure}

I have not found interesting evidence of the continuation of Sheets~A or~B in the 497 MCXC clusters in the redshift bin $0.085<z<0.17$, or the 209 XMM X-ray cluster detections (Koulouridis, Clerc, Sadibekova, et al. 2021), and there are few 2MRS galaxies at this distance. This makes it interesting to estimate how far detection of Sheet~A$_3$ extends into the third redshift bin. The minimum and maximum radial distances from the origin in the $j=-1$ cell are 182.2 and 298.5~Mpc. The solid histograms in Figure~\ref{fig:disthist} (black for 2MRS galaxies, red for MCXC clusters) show the distributions of radial distances, from redshifts, of the objects that are in the disk cell $j=-1$ when the tube axis is in the direction in equation~(\ref{eq:123}). The dotted histograms show the counts of these objects as a function of radial distance in cell $j=-1$ in the average over statistically isotropic random directions. The fluctuations in these dotted histograms are real, determined by the positions of the objects. We see that at distances greater than 230~Mpc there are only marginally more objects in the cell oriented along the continuation of Sheet~A than would expected in the absence of this feature. The counts above average at distances less than 230 are modest but systematically larger than random, as we see in Figure~\ref{fig:hist123}. I conclude that Sheet~A$_3$ is reasonably well detected to about 230~Mpc distance, or redshift $z\sim 0.05$. 

\section{Discussion}\label{sec:discussion}

In the standard and well-tested $\Lambda$CDM theory the primeval mass distribution is a stationary random Gaussian process with a  close to scale-invariant power spectrum without the patterns under discussion here. The theory follows in a natural way from the cosmological inflation picture for the evolution of the very early universe, and the theory fits precision measurements of the anisotropy of the cosmic microwave background and a considerable network of other critical and well-checked tests that persuasively establish the theory as a useful approximation to reality (e.g. Peebles, Page, and Partridge 2009). But dark matter and dark energy were  put in the theory as the simplest way to get agreement with the empirical evidence we had decades ago (Peebles 1982, 1984), and although the Gaussian initial conditions are a natural consequence of inflation they are not a prediction. These considerations are meant to remind the reader of two points. First, evidence of phenomena that would require adjusting the $\Lambda$CDM theory must be treated with caution because the theory passes demanding tests. But second, the search for evidence of phenomena that would require adjusting the theory is essential because the theory is incomplete and might be improved by following hints from empirical evidence.

\subsection{The Evidence}\label{sec:evidence}

The phenomena discussed here are the signatures of large-scale thin and flat sheet-like patterns in cosmic structure. The method of detection of these phenomena can equally well be applied to realizations of theoretical predictions of the spatial distribution of galaxies or clusters of galaxies. This offers the prospect of an interesting and maybe critical comparison of theory and observation that might  lead us to a still better theory. 

I have treated with caution the evidence of phenomena that seems to require adjusting the successful $\Lambda$CDM theory. To begin, at redshifts $z<0.021$ Sheets~A and~B are detected in catalogues obtained in the three quite different ways reviewed in Section~\ref{sec:reproducibility}. This makes a good case that the detection is not some exceedingly subtle systematic error in the data. 

Even more compelling evidence of reality of Sheet~A is the continuity of detection of the patterns in data that are disjoint (or very close to it in the Nearby Universe). The Nearby Universe and Local Supercluster are known to be flat; it was the motivation for this  study. The probe used to detect flat patterns applied to this region at distances $\lap 7$~Mpc gives the indication of its  flatness in Figure~\ref{fig:NU}. By this measure the Nearby Universe is tilted $5^\circ$ from the de Vaucouleurs supergalactic pole. Figures~\ref{fig:survey11} to \ref{fig:PSCz} show evidence of the presence of Sheet~A$_1$ in the redshift bin $z<0.021$, the Local Universe. The normal to this sheet is tilted $11^\circ$ from the supergalactic pole and $15^\circ$ from the sheet of the Nearby Universe. The Nearby Universe sheet and Sheet~A$_1$ both contain the origin, our position. We can take the continuity in position and near continuity in direction to mean the Nearby Universe is a moderately bent part of Sheet~A$_1$ in the Local Universe. This is an important example of continuity of detection in different ranges of length scales. 

Detection of Sheet~A$_2$ at  $0.021<z<0.042$ is shown in Figure~\ref{fig:survey12}. It is tilted $8^\circ$ from Sheet~A$_1$, $21^\circ$ from the Nearby Universe, and $18^\circ$ from the supergalactic pole. The disk cell used here has a hole in its center, to avoid the galaxies used in the first redshift bin, but the centerplane of the cell passes through the origin (Fig.~\ref{fig:d2}), along with the Local Supercluster. The detection of this feature about where it would be expected in position and orientation amounts to a confirmation of a prediction. This is substantial evidence of the physical reality of Sheet~A. 

At the third redshift bin, $0.042<z<0.085$, the evidence that Sheet~A$_3$ is detected is seen in Figures~\ref{fig:survey13} and~\ref{fig:survey123}. Sheet~A$_3$ is tilted $5^\circ$ from Sheet~A$_2$, $3^\circ$ from Sheet~A$_1$,$17^\circ$ from the plane of the Nearby Universe, and $13^\circ$ from the supergalactic pole. The centerplane of the cell that contains Sheet~A$_3$  is 20~Mpc from the origin. This is modest compared to the dimensions of the disk cell, 20~Mpc thick by 600~Mpc diameter, but the offset in distance from the pole is real. Maybe it signifies a bend or warp in the sheet tracing back to the sheet of the Nearby Universe, or maybe something more complicated. But the postulate of a warp is in the philosophy of tolerating modest bends in the segments of redshift that detect Sheet~A. 

Sheet~B$_1$ in the first redshift bin looks similar to Sheet~A$_1$. This is seen in the counts in disk cells as functions of distance from the cells containing the sheets (Fig.~\ref{fig:2sheets}), the distributions of galaxy morphologies (Fig.~\ref{fig:morphologyratios}), the maps (Fig.~\ref{fig:maps}), and the sensitivity to tilts of the disk cell (Fig.~\ref{fig:shapes}). Sheet~B$_2$ is in the interval of distance from the origin $-25<h<-15$~Mpc, consistent with the interval $-17.5<h<-12.5$~Mpc in the first redshift bin. The normal of Sheet~B$_2$ is $15^\circ$ from Sheet~B$_1$. This signature of a modestly bent continuation of Sheet~B in two disjoint intervals of redshift is significant evidence of another thin close to flat sheet in cosmic structure. It might be expected that the sheet of the Local Supercluster cannot be  unique, but the evidence of another specific sheet is important. 

It is curious that the 51 clusters in the first redshift bin yield clear signatures of Sheets~A$_1$ and~B$_1$ (Fig.~\ref{fig:CLs}), and the 351 clusters in the third bin give a clear signature of Sheet~A$_3$ (Figs.~\ref{fig:survey123} and~\ref{fig:hist123}), but the 139 clusters in the second bin yield at best a modest suggestion of detection (in the lower plot in Fig.~\ref{fig:d2}). But there are reasonably clear signatures of Sheets~A and~B in the distribution of galaxies in the second bin (Figs.~\ref{fig:survey12} and~\ref{fig:d2}).

To summarize, the near continuity in directions and distances of the normals of the sheet-like patterns in three disjoint redshift bins, and in the close to disjoint Nearby Universe, is serious evidence of reality of Sheet~A, the Extended Local Supercluster, which extends to redshift $z\sim 0.05$. 

\subsection{Consequences}

The probe that establishes this case for reality of flat sheets could readily be applied to spatial distributions of simulated galaxies, or of dark matter halos characteristic of galaxies or clusters of galaxies. Of greatest interest are positions established by numerical simulations of the growth of cosmic structure from the initial conditions and physics of the present standard $\Lambda$CDM theory.  A likely first step would be a check for occurrence in numerical simulations of the flat Nearby Universe, the part of the Local Supercluster shown in Figure~(\ref{fig:NU}). Neuzil, Mansfield, and Kravtsov (2020) remark on the ``flat sheetlike shape'' of the Local Supercluster, and find that the fit of an ellipsoid to the distribution of galaxies closer than 8~Mpc has ratio of largest to smallest axes $a/c\sim 6$. This is in the spirit of Figure~\ref{fig:NU}, though not as flat as suggested by the figure. Neuzil et al. (2020) find that their measure of flattening is ``not exceedingly rare in the $\Lambda$CDM model.'' The pronounced concentration of galaxies in the central sheet 0.5~Mpc thick and 15~Mpc across in the Nearby Universe  intuitively seems more demanding than the Neuzil el al. (2020) criterion, but a check certainly would be of great interest. 

I have not found discussions of whether the  $\Lambda$CDM theory might be challenged by the considerable extension of the plane of the Local Supercluster found by Einasto et al.~(1983), Tully (1986), and Shaver (1990). The extension has been more precisely characterized here, and again a check for the presence of close to flat patterns similar to Sheets~A and~B in simulations of cosmic structure certainly would be interesting.

If the patterns of flat thin sheets prove to be real and not predicted by the $\Lambda$CDM theory then the theory will remain a very useful approximation, but it will have been established that we have an interesting hint to an even better theory to be found. 

If as seems intuitively likely the signatures of flat and extended sheets presented here are real and are not predicted by the standard $\Lambda$CDM cosmology then thoughts will turn to the passages of long nearly straight cosmic strings that produced near-planar density concentrations (e.g. Shellard, Brandenberger, Kaiser, and Turok 1987; Vachaspati and Vilenkin 1991; da Cunha, Harnois-Deraps, Brandenberger, Amara, and Refregier 2018). Could cosmic strings be straight enough to produce extended flat sheets and less massive enough to agree with the constraint from data assembled by the International Pulsar Timing Array (Falxa M., Babak S., Baker P.~T. et al., 2023)? If so would cosmic string perturbations to the galaxies and clusters in sheets agree with the observations? But the immediately pressing question is whether cosmic strings capable of producing the flat patterns discussed here would have observable effects on the angular distribution of the CMB. One would look for the distinctive line discontinuities Kaiser and Stebbins (1984) and Gott (1985) predicted. The effect on the CMB might be detected by N-point  statistics (e.g. Planck Collaboration et al. 2014 and references therein), or maybe by a probe sensitive to the expected nature of perturbations to the CMB temperature caused by moving strings. Danos and Brandenberger (2010) discussed a probe of temperature gradients. Another possibility is an adaption of the filter used in this study. 

If cosmic strings fail we will need a new idea. Situations of this kind have led to many advances in the development of natural science.

\section*{acknowledgements}\label{sec:acknowledgements}

I have been suggesting that the curious extent of the plane of the Local Supercluster might be teaching us something of value for more than a quarter century, beginning with the plane traced by clusters in Peebles~(1993, Fig. 3.7c), then following Shaver's (1990, Fig.~1; 1991, Fig.~5) elegant demonstration of the alignment of  radio galaxies the argument that we ought to be thinking about this in Peebles (1999) and four  later publications. My first step to follow this advice is in Peebles (2022). In the present paper I have taken the Extended Local Supercluster to be the prototype for the search for thin flat patterns in cosmic structure. I remember with pleasure discussions of this and more with Peter Shaver. I am grateful for information and advice on this paper from Miguel Aragon, Robert Brandenberger, Roger Clowes, Maret Einasto, Andrey Kravtsov, Biswajit Pandey, Daniel Pomar\`ede, Joe Silk, Brent Tully, and Yen-Ting Lin. I am grateful to Princeton University for giving me the space to do this research. I have not received any financial support for this research, and I am not complaining.

\section*{Data Availability}

This research made use of the data compiled by the  High Energy Astrophysics Science Archive Research Center (HEASARC), which is a service of the Astrophysics Science Division at NASA/GSFC; the SIMBAD and  VizieR catalogue access tool, CDS, Strasbourg, France; the Two Micron All Sky Survey, which is a joint project of the University of Massachusetts and the Infrared Processing and Analysis Center/California Institute of Technology, funded by the National Aeronautics and Space Administration and the National Science Foundation; the NASA Astrophysics Data System Bibliographic Services; and the NASA/IPAC Extragalactic Database. I have not  generated any new data.

\label{lastpage}
\end{document}